\documentclass{jfm}

\usepackage{graphicx}
\usepackage{natbib}

\ifCUPmtlplainloaded \else
  \checkfont{eurm10}
  \iffontfound
    \IfFileExists{upmath.sty}
      {\typeout{^^JFound AMS Euler Roman fonts on the system,
                   using the 'upmath' package.^^J}%
       \usepackage{upmath}}
      {\typeout{^^JFound AMS Euler Roman fonts on the system, but you
                   dont seem to have the}%
       \typeout{'upmath' package installed. JFM.cls can take advantage
                 of these fonts,^^Jif you use 'upmath' package.^^J}%
      }
  \else
  \fi
\fi

\ifCUPmtlplainloaded \else
  \checkfont{msam10}
  \iffontfound
    \IfFileExists{amssymb.sty}
      {\typeout{^^JFound AMS Symbol fonts on the system, using the
                'amssymb' package.^^J}%
       \usepackage{amssymb}%
       \let\le=\leqslant  
       \let\ge=\geqslant  
      }{}
  \fi
\fi

\ifCUPmtlplainloaded \else
  \IfFileExists{amsbsy.sty}
    {\typeout{^^JFound the 'amsbsy' package on the system, using it.^^J}%
     \usepackage{amsbsy}}
    {}
\fi

\newsavebox{\astrutbox}
\sbox{\astrutbox}{\rule[-5pt]{0pt}{20pt}}

\title[C. H. Gibson]{Falsification of dark energy by fluid mechanics}

\author[C. H. Gibson]%
{C\ls A\ls R\ls L\ns H.\ns G\ls I\ls B\ls S\ls O\ls N$^1$%
  \thanks{Email address for correspondence: cgibson@ucsd.edu}}

\affiliation{$^1$Departments of MAE and SIO, Center for Astrophysics and Space Sciences, University of California at San Diego,
La Jolla, CA 92093-0411, USA\\}

\pubyear{2010}
\volume{650}
\pagerange{119--126}
\date{?; revised ?; accepted ?. - To be entered by editorial office}
\begin{document}

\maketitle

\begin{abstract}
The 2011 Nobel Prize in Physics was awarded for the discovery of accelerating supernovae 
dimness, suggesting a remarkable change in the expansion rate of the Universe from a decrease since the big bang to 
an increase, driven by anti-gravity forces of a mysterious dark energy
material comprising 70$\%$ of the Universe mass-energy. Fluid mechanical considerations
falsify both the accelerating expansion and dark energy concepts.  Kinematic viscosity is
neglected in current standard models of self-gravitational structure formation, which rely on
cold dark matter CDM condensations and clusterings that are also falsified by fluid mechanics.  Weakly
collisional CDM particles do not condense but  diffuse away.  Photon viscosity predicts
superclustervoid fragmentation early in the plasma epoch and protogalaxies at the end.  At the plasma-gas 
transition, the plasma fragments into Earth-mass gas planets in trillion planet clumps (proto-globular-star-cluster PGCs).
The hydrogen planets freeze to form the dark matter of galaxies and merge to form their stars.  Dark energy is a systematic dimming error
for Supernovae Ia caused by dark matter planets near hot white dwarf stars at the Chandrasekhar carbon limit.  Evaporated planet atmospheres may or may
not scatter light from the events depending on the line of sight.
\end{abstract}

\begin{keywords}
Authors should not enter keywords on the manuscript, as these must be chosen by the author during the online submission process and will then be added during the typesetting process (see http://journals.cambridge.org/data/\linebreak[3]relatedlink/jfm-\linebreak[3]keywords.pdf for the full list)
\end{keywords}

\section{Introduction}
For any physical theory to be useful it must be falsifiable; that is, it must be subject to testing by observations that could render it false.  The 2011 Nobel Prize in Physics has been awarded to Perlmutter, Reiss and Schmidt for observations of distant supernovae showing the rate of expansion of the Universe ceased to decrease after the big bang and is now increasing.  The purpose of the present paper is to show this remarkable interpretation of supernovae dimness is falsified by dark matter planets in clumps predicted from basic fluid mechanics. The planets in clumps are supported by observations that confirm the falsification. An alternative fluid mechanically based cosmology is proposed.

The standard (concordance) model of cosmology ($CC$) is $\Lambda CDMHC$, where $\Lambda$ represents the proposed dark energy material driving the accelerating expansion rate of the Universe by anti-gravitational forces.  $CDM$ is Cold Dark Matter, an unknown and at least equally massive (and equally questionable) cosmological material that condenses by hierarchical clustering $HC$.   $CC$ cosmology and $CDM$ dynamics are  discussed in the following $\S{2}$ on Theory.  The fatal flaw of the $CC$ model is its assumption that fluid mechanics can be ignored assuming ideal collisionless potential flow.  

Numerous attempts have been made to include basic fluid mechanical improvements to the $CC$ model starting with \cite{gib96}; that is, to include turbulence, kinematic viscosity, diffusivity, stratified fossil turbulence and turbulent mixing.  An initial finding was that the dark matter of galaxies must be Earth-mass primordial gas planets in dense Jeans-mass clumps.  This prediction of \cite{gib96} was immediately confirmed by \cite{schild96} from his 15 year study of quasar microlensing by a galaxy on the line of sight.   Schild inferred from the twinkling frequency and brightness of the quasar images that the missing mass and dominant point mass objects of the galaxy must be small planets  in clumps, not stars. The improved $CC$ cosmology is termed hydro-gravitational-dynamics, or $HGD$, cosmology.  $HGD$ predicts $3 \times10^{7}$ planets per star in a galaxy, as observed, \cite{schild96}: not 8. 

A hot big bang event at Planck scales is overwhelmingly accepted as the simplest explanation of observations from modern telescopes in the context of Einstein's general relativity theory and quantum mechanics, \cite{Peacock}.  However, gravitational accelerations of $10^{51}$ g must be overcome at Planck length scales $10^{-35}$ m, \cite{gib05}.  Gravitational pressures of $10^{113}$ Pa are required, and are matched by negative stresses of  big bang turbulence vortex stretching, \cite{gsw10}. Spin from big bang turbulence fossils, \cite{gib04}, is preserved at $10^{25}$ m length scales, \cite{sg11}, $10\%$ of the present horizon scale $L_H$.  

$CDM$ hierarchical clustering $HC$ has been recently falsified by \cite{Kroupa10} from observations showing the required numbers of small clusters of stars (Local Groups) do not exist in the Milky Way galaxy.  This finding falsifies dark energy $\Lambda$ by removing its theoretical basis.  Small dark matter gas planets are also ruled out by $CDMHC$ cosmology.  Every small gas planet detected falsifies $\Lambda$. $HGD$ cosmology predicts a rapid $\sim t^{3}$ formation of galactic central black holes, Nieuwenhuizen (2011).  Quasars observed at redshifts $z \ge 10$ therefore falsify both $CDMHC$ and $\Lambda$ from fluid mechanics.
 
\section{Theory}
Concordance $CC$ cosmology rejects basic fluid mechanics as unnecessary.  Flows are assumed to be ideal, resulting from gradients in a velocity potential.  Gravitational structure formation is based on the Jeans analysis of 1902. Table 1 presents the alternative $HGD$ scenario. Proto-galaxies form at the end of the plasma epoch with Nomura scale ($L_N$) and morphology and evolve  on fossil vorticity turbulence vortex lines stretched by the expansion of the universe, \cite{gs10a} and \cite{gs10b}.

\cite{jeans02} reduced the Navier Stokes equations to acoustics by neglecting viscosity (Euler's equation), neglecting density (Jeans swindle), neglecting non-linearity by the application of linear perturbation stability analysis (no turbulence), and neglecting diffusion of weakly interacting massive particles (no neutrinos).  Self gravitational structure formation is permitted only for length scales $L_H \ge L \ge L_J = V_S \tau_g$, where $V_S$ is the speed of sound, $ \tau_g = (\rho G)^{-1/2}$ is the gravitational free fall time, $\rho$ is density, $G$ is Newton's gravitational constant, the horizon scale  $L_H = ct$ is the scale of causal connection, $c$ is light speed, and $t$ is time since the big bang.

Without fluid mechanics, cold dark matter is needed to explain structure formation during the plasma epoch $10^{11} \le t \le 10^{13}$ seconds because $L_J \ge L_H$ for the plasma during this period.  With fluid mechanics, it is only necessary to recognize that the low densities and high temperatures of the early $H-He^4$ plasma and gas Universe give large kinematic viscosities $\nu$ values that dominate self-gravitational structure formation, contrary to \cite{jeans02} and all $CDM$ cosmology models. In the plasma epoch photons transport momentum by elastic collisions with electrons (Thomson scattering) that communicate it to the ions. Protosupercluster fragmentation occurs at $10^{12}$ s when $L_H \ge L_{SV}$.

Most of the dark matter of the Universe is non-baryonic (not composed of protons and neutrons).  Whatever its particles are, they must be nearly collisionless, like neutrinos, or they would have been detected.  It was assumed by $CC$ cosmology that somehow non-baryonic dark matter NBDM material would have a smaller Jeans scale if it were cold, so it could condense by the Jeans 1902 criterion.  These small condensate seeds would then clump together ($HC$) to make massive halos into which the baryonic dark matter would fall and condense as stars, galaxies and galaxy clusters (in this order).    Such assumptions are fluid mechanically untenable. 

Consider a clump of perfectly cold non-baryonic dark matter (NBDM), or a cluster of smaller clumps termed a halo.    The initially motionless particles immediately start to move toward the center of gravity of the halo.  No matter how small the collision cross section of $CDM$ particles is assumed to be, a density will be reached (in a free fall time) large enough for the collisionless assumption to fail, so the halo simply diffuses away.  The size of $CDM$ halos is found by matching the diffusion velocity to the gravitational velocity at the Schwarz diffusive length scale $L_{SD} = (D^2 \rho G)^{1/4}$, where $D$ is the diffusivity.  From fluid mechanics, $D = (n/\sigma)v_p$, $n$ is the particle number density, $\sigma$ is the particle collision cross section and $v_p$ is the particle velocity.  No $CDM$ structure formation is possible during the plasma epoch because $L_{SD} \ge L_H$ during this period, \cite{gib96}.  

Viscous forces balance self-gravitational forces at the Schwarz viscous scale $L_{SV} = ({\gamma \nu}/{\rho G})^{1/2}$, where $\gamma$ is the rate of strain and $\nu$ is the kinematic viscosity.  Viscous forces were small in the beginning instant of the big bang event $10^{-43}$ s because the Planck  length scale $10^{-35}$ m for collisions of Planck particles was the only one available at Planck temperatures $10^{32}$ K. Viscous and inertial vortex forces matched at the Kolmogorov scale, \cite{gib05}, so the Schwarz turbulence scale $L_{ST} = (\varepsilon / [\rho G)]^{3/2})^{1/2}$ remained larger than the Planck scale  and Schwarz viscous $L_{SV}$ scale during the big bang turbulence event until quenched by gluon-viscosity at the strong force freeze out temperature $10^{28}$ K.  Viscous dissipation rates $\varepsilon$ of $10^{60}$ W homogenize small scales of the turbulence.  Negative gluon-viscous stresses power anti-gravitational inflation of the fossilizing turbulent fireball to meter scales preserving Planck density, \cite{gib04}.

  A $10^{25}$ m fossil big bang turbulence vortex line persists as the axis of evil (a preferred direction on the sky), \cite{sg11}.  Such a preferred direction on large scales falsifies the intrinsically scalar aspect of the dark energy concept $\Lambda$.  Turbulence effects indicated on the cosmic microwave background (studies of Sreenivasan and Bershadski) have been discussed, fig. 5, \cite{gib10}.

\begin{table}
  \begin{center}
\def~{\hphantom{0}}
  \begin{tabular}{lccc}
      $t, seconds$  &  $event$  \\[3pt]
       $10^{11}$   & mass exceeds energy, plasma epoch begins, $L_H \le L_{SV}$\\
      $10^{12}$   & protosupercluster fragmentation occurs when $L_H \ge L_{SV}$, $10^{5}$ K\\
       $10^{13}$   & $L_N = 10^{20}$ m plasma protogalaxy fragmentation on vortex lines\\
        $10^{13}$   & $\nu_{plasma} = 10^{26} m^2~s^{-1}$ (photon viscosity), $\nu_{gas} = 10^{13} m^2~s^{-1}$\\
        $10^{13}$   & gas epoch begins, 3000 K, $\rho = \rho_{0} = 4 \times 10^{-17} kg ~m^{-3}$\\
        $10^{13} + 10^{12}$   & planets fragment in Jeans mass clumps (protoglobularstarclusters)\\
        $10^{13} + 10^{12}$   & first stars appear as gas planets merge (no $10^{16}$ s dark age)\\
         $10^{13} + 10^{12}$   & first chemicals form in stars and supernovae, exploding stars seed planets\\
         $10^{13} + 10^{12}$   & iron cores condense as planets merge, oxides are reduced to metals and water\\
         $10^{14} $   &oceans condense at critical water temperature 647 K, life begins everywhere\\
         
  \end{tabular}
  \caption{HGD cosmology plasma to gas transition events $\to$ the biological big bang}
  \label{tab:kd}
  \end{center}
\end{table}

\begin{figure}
  \centerline{\includegraphics{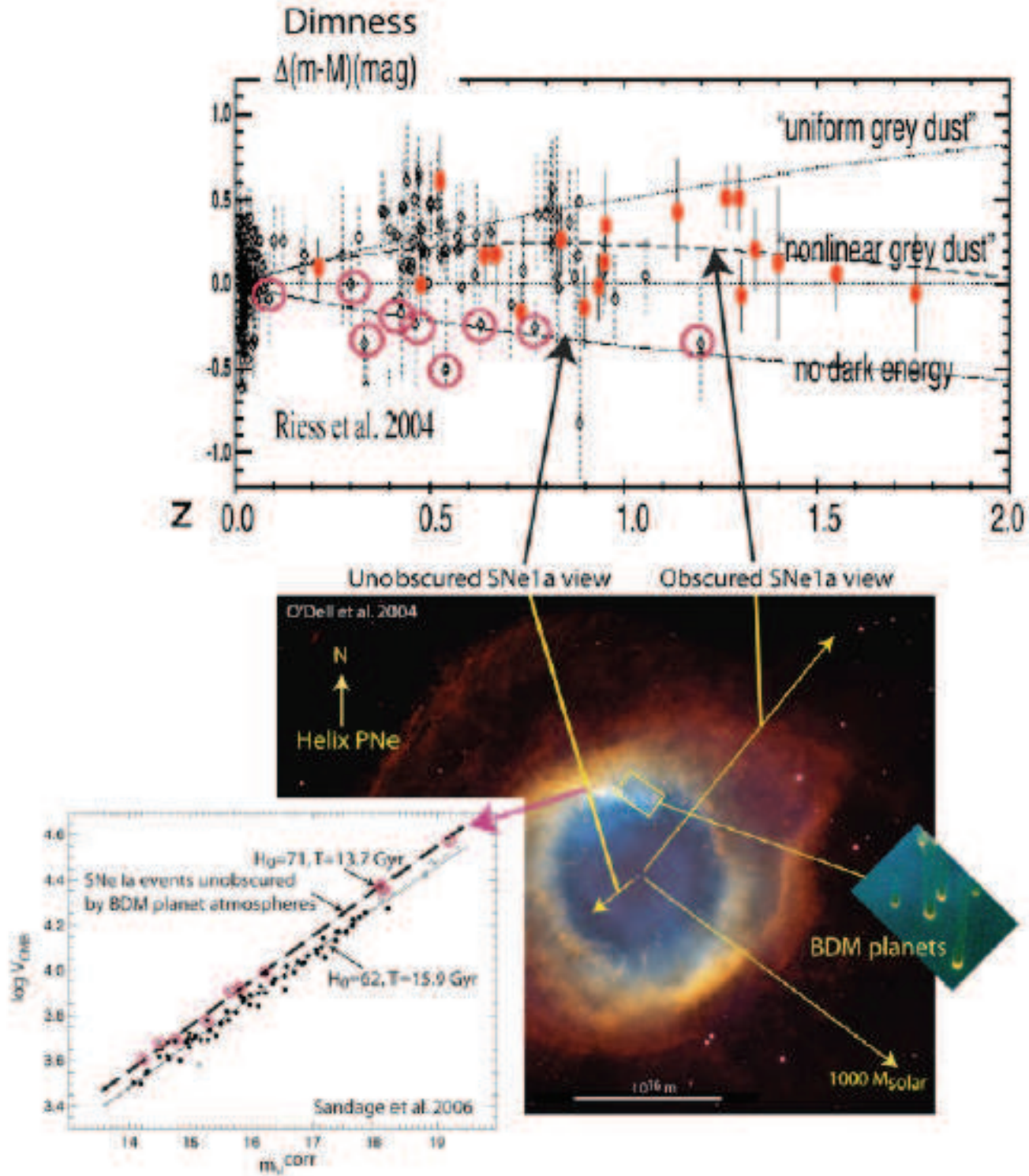}}% Images in 100% size
  \caption{Dark energy is a systematic dimming error due to baryonic dark matter $BDM$ planet atmospheres surrounding dying stars such as the White Dwarf in Helix (fig. 8, \cite{gib10}).}
\label{fig:ka}
\end{figure}

\begin{figure}
%  \centerline{\includegraphics[height=7cm,width=13cm]{modes.eps}}
  \centerline{\includegraphics{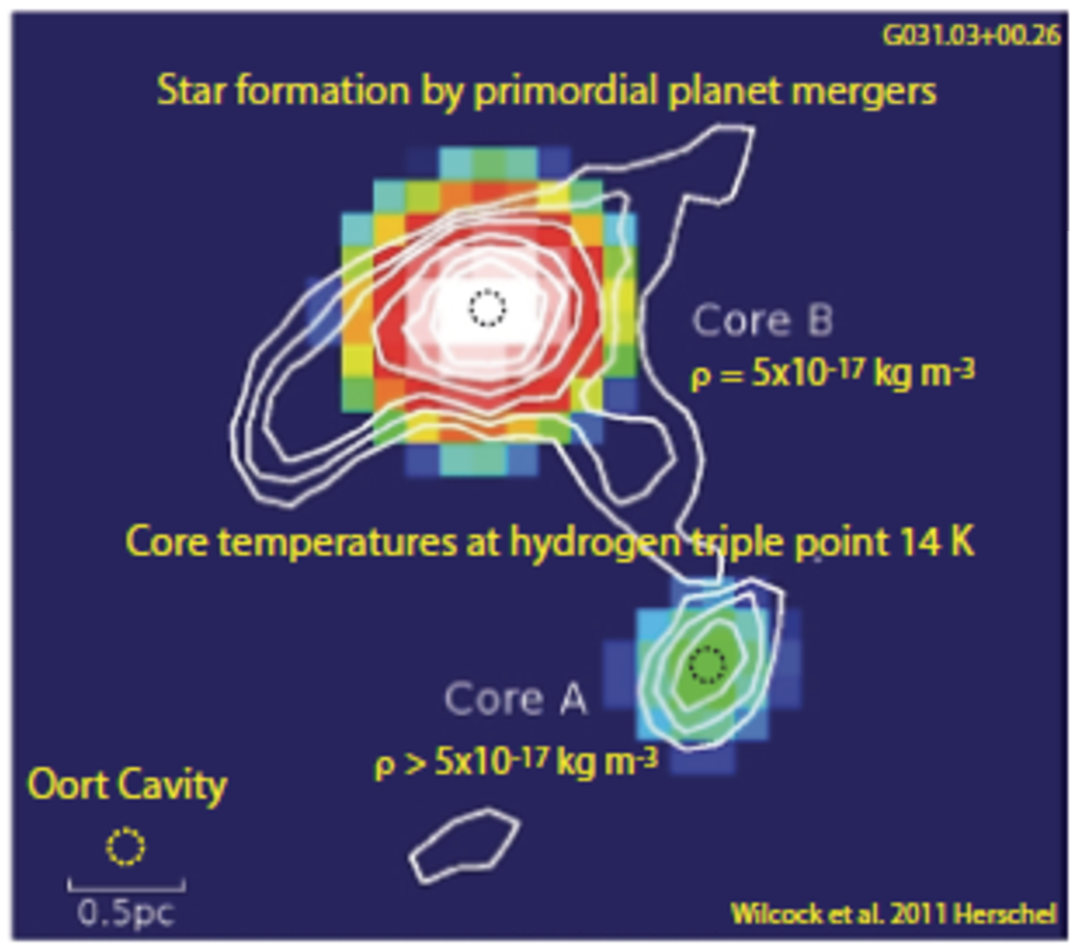}}
  \caption{Herschel space telescope study of dense cores, \cite{wilcock11}, showing evidence of star formation by frozen primordial planet mergers.  Both Core A and Core B have densities matching $\rho_0 = 4 \times 10^{-17} kg~m^{-3}$ existing at the time of first fragmentation $10^{12}$ seconds, and merging temperatures expected for evaporating planets at the triple point of hydrogen  14 K. }
\label{fig:kd}
\end{figure}

\section{Observations}

Figure 1 illustrates falsification of dark energy by the fluid mechanics of $HGD$ cosmology.  Hubble Space Telescope optical observations of the Helix planetary nebula (bottom) are compared to Supernovae Ia dimness observations (top) as a function of redshift $z$ (\cite{gib10}, fig 8).  Further discussion is found in the reference. 

The dark energy proposition is falsified because it does not take into consideration the possibility that the atmospheres of evaporated dark matter planets (Table 1) may randomly dim or not dim the light of Sne Ia  events depending on the line of sight.

Error bars on the dimness data points (top fig 1) are smaller than the scatter, and cannot be explained as "uniform grey dust". Solid ovals support a "nonlinear grey dust" explanation of a systematic dimming by planet atmospheres on the line of sight, and open circles support a "no dark energy" interpretation where the Sne Ia events are unobscured.  The same systematic error is seen (bottom left) in  Sne Ia estimates of the age of the Universe, which also falsify Dark Energy and confirm $HGD$ cosmology from fluid mechanics.
 
Figure 2 illustrates dark energy falsification by fluid mechanics from the Herschel space observatory, \cite{wilcock11}.  The wide range of frequencies permit detection of low temperatures and high densities of dense core regions with star formation.  The core densities closely match the high density of first fragmentation $\rho_{0}$ expected from $HGD$ cosmology (Table 1).  Core temperatures 14 K match the triple point temperature of hydrogen (13.8 K), attributed to evaporating planets merging to form stars.  Black dots circling the core at $10^{16}$ m represent large gas planets evaporating at Oort cloud distances.  Such planets are shown (bottom right insert fig. 1) at this distance.  Their size  $\ge 10^{13}$ m indicates previous dark matter planet mergers. 

\section{Conclusions}

The accelerating expansion rate of the Universe and Dark Energy concepts rewarded by the 2011 Nobel Prize in Physics are falsified by fluid mechanics. The standard (Concordance Cosmology)  $\Lambda CDMHC$ model of cosmology must be replaced by $HGD$ cosmology.

\end{document}